\documentclass[reprint,english,preprintnumbers,amsmath,aps,prl]{revtex4-2}
 \usepackage{graphicx}
\usepackage{dsfont}
\usepackage{color}
\usepackage{multirow} 
\usepackage{charter} 
\usepackage[charter]{mathdesign}   
\definecolor{refcol}{RGB}{0,100,205} 
\usepackage{microtype}
\usepackage[colorlinks,linkcolor=refcol,citecolor=refcol,urlcolor=refcol]{hyperref}
\begin{document}

\title{Signatures of deuteron synthesis on the modified combinants in nuclear collisions}
\author{Rahul R Nair}
\email{rrn.phy@gmail.com}
\author{Grzegorz Wilk} 
\email{grzegorz.wilk@ncbj.gov.pl}
\affiliation{National Centre For Nuclear Research, Pasteura 7, Warsaw 02-093, Poland.} 
\author{Zbigniew Włodarczyk}
\email{zbigniew.wlodarczyk@ujk.edu.pl}
\affiliation{Institute of Physics, Jan Kochanowski University, 25-406 Kielce, Poland.}
  
\begin{abstract}
The mechanism of deuteron production in heavy-ion collisions is shown to have a significant impact on the shape of the modified combinants of their multiplicity distribution. In light of this observation, an experimental study is proposed to tackle the long-standing problem of nuclei synthesis in hadronic and nuclear collisions. The proposed approach has the potential to provide stringent constraints on the models of deuteron synthesis.
\end{abstract}

\maketitle


{\bf Introduction --}\label{sec:intro}
The mechanism of deuteron production in high-energy nuclear and hadronic collisions remains unresolved despite extensive studies spanning several decades, from AGS to the LHC era \cite{PhysRevC.60.064901,PhysRevC.61.064908,PhysRevC.61.044906,PhysRevC.65.034907, KABANA1999370, Bearden2002,200022, PhysRevC.94.044906, PhysRevC.69.024902, PhysRevC.85.044913, PhysRevLett.99.052301, PhysRevC.79.034909,https://doi.org/10.48550/arxiv.0909.0566,Agakishiev2011,PhysRevC.93.024917,PhysRevC.102.055203,Acharya2020}. The thermal and coalescence models are the two dominant theories for explaining deuteron synthesis in such collisions. These models predict the experimental yield of deuterons in a variety of collision systems and energies. In the thermal model, the ratio of particle yield is determined solely by the temperature of the system and the baryon chemical potential at chemical freeze-out\cite{ANDRONIC2011203, PhysRevC.84.054916, doi:10.1142/9789812795533_0008}. The coalescence model posits that nucleons emitted from a fireball with small relative momenta form light nuclei due to attractive nuclear forces \cite{PhysRev.129.854, PhysRev.129.836}. There exist various scenarios for light nuclei production via coalescence in the literature \cite{s2020,s2017, PhysRevC.99.054905}. It provides a microscopic explanation for light nuclei synthesis, with its final yield dependent on various parameters of the model. Despite being studied for decades, a comprehensive understanding of the production of nuclei remains an open question in nuclear physics.
\par
On the other hand, the modified combinants, denoted by $C_j$, of the particle multiplicity distribution $P(n)$ in high-energy collisions have been demonstrated to display certain intriguing features that can provide insights into the production process of the particles \cite{Wilk_2017,doi:10.1142/S0217751X2150072X, Ang2020, PhysRevD.99.094045,doi:10.1142/S0217751X18300089}. The multiplicity distributions can be characterized by their generating functions $G(z)$ as 
\begin{equation}
    G(z) = \sum^{\infty}_{n=0} P(n)z^n
\end{equation}
or through the recursive formula 
\begin{equation}\label{eq2}
    (n+1)P(n+1) = g(n)P(n)
\end{equation}
where the function $g(n)$ determines the algebraic structure of $P(n)$. The form of $g(n)$ can be selected to match the experimental measurement of the multiplicity. By considering that $g(n)$ should reflect the interconnections between multiplicity $n$ and all lower multiplicities, Eq.\ref{eq2} can be written as follows
\begin{equation}\label{eq3}
    (n+1)P(n+1) = \left<n\right>\sum_{j=0}^{n}C_jP(n-j)
\end{equation}
where the coefficients $C_j$ are the modified combinants. In this study, the method of modified combinants developed in \cite{Wilk_2017,doi:10.1142/S0217751X2150072X, Ang2020, PhysRevD.99.094045,doi:10.1142/S0217751X18300089} is applied to investigate deuteron synthesis in heavy-ion collisions. The following section details the construction of modified combinants and revisits the relevant expressions used in subsequent calculations.

\vspace{5pt}
{\bf Modified combinants --}\label{sec:Mod}
\begin{figure*}[hbt!]
\includegraphics[width=\textwidth,height=0.5\textwidth]{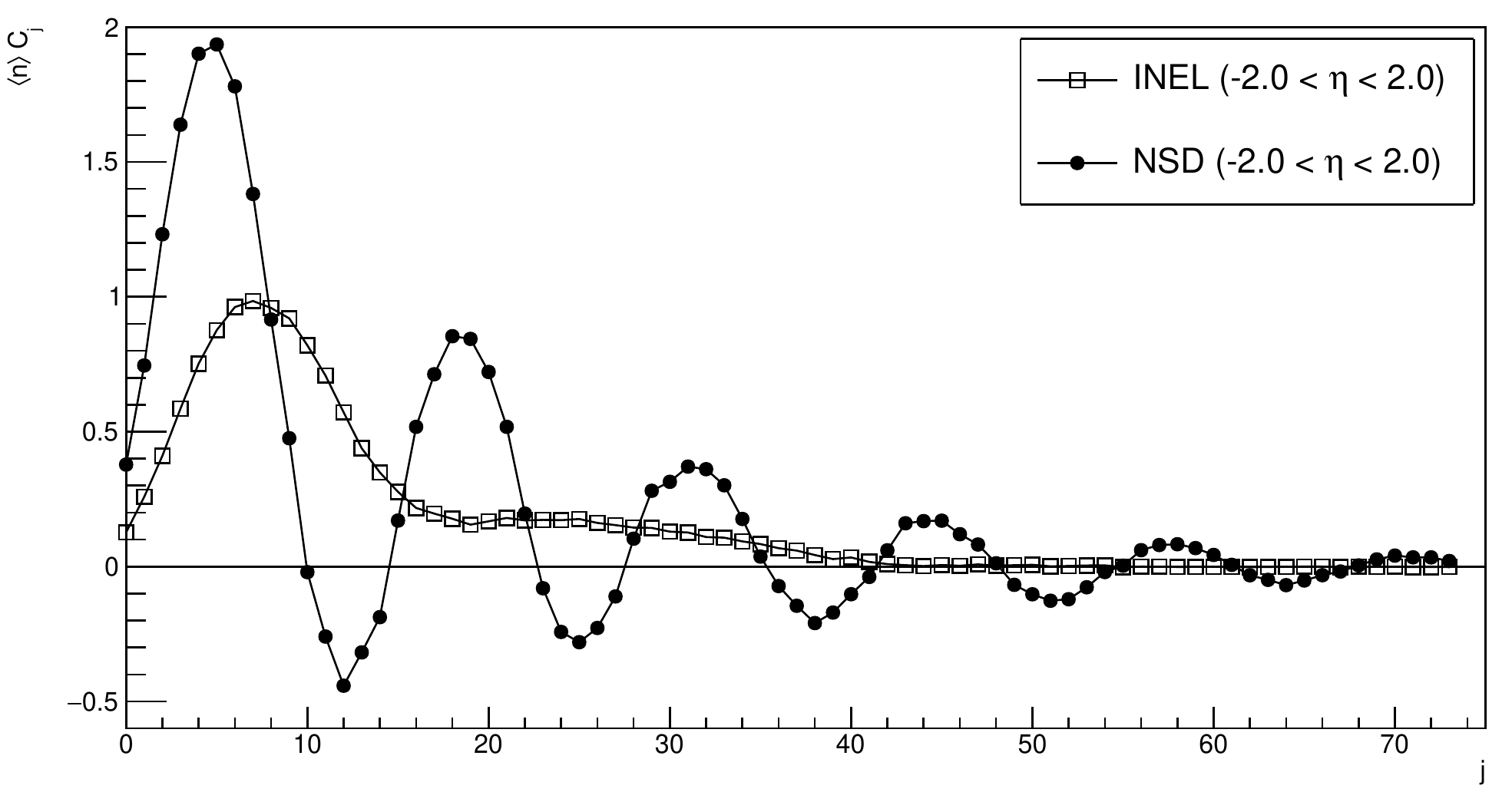}
 \caption{Modified combinants of charged particle multiplicity distribution for NSD and INEL pp collisions in ALICE at $\sqrt{s_{NN}} = $ 900 GeV in the pseudorapidity range of $|\eta|< 2.0$. Data is taken from \cite{ALICE:2017pcy}. The solid lines are to lead the eye.}
 \label{fig1}
\end{figure*}
 The multiplicity distribution in high-energy collisions is a widely used observable and is accessible from many collider experiments \cite{Botet:2002gj, doi:10.1142/5805}. The probability of producing $n$ particles in a collision, denoted by $P(n)$, is an important distribution that can be used in many analyses. Recently, many studies have shown that even more information about the particle production process can be obtained from the measured multiplicity  distributions P(n) by examining their modified combinants. By reverting the recurrence relation in Eq.~\ref{eq3}, an expression for the modified combinants can be obtained as
\begin{equation}
\label{Comb}
    \left<n\right>C_j = (j+1) \frac{P(j+1)}{P(0)}  - \left<n\right>\sum_{i=0}^{j-1}C_i \frac{P(j-i)}{P(0)} 
\end{equation}
where $\left<n\right>$ is the mean of the multiplicity distribution. They can also be defined in terms of the generating function $G(z)$ as
\begin{equation}\label{eq5}
    P(n) = \frac{1}{n!}\frac{d^nG(z)}{dz^n}\Big|_{z=0}
\end{equation}
and 
\begin{equation}
    C_j = \frac{1}{j!}\frac{d^{j+1}}{dz^{j+1}}\ln G(z) \Big|_{z=0}
\end{equation}
The cumulant factorial moments $K_q$ are related to the modified combinants as follows
\begin{equation}
  K_q = \sum_{j=q}^{\infty}  \frac{(j-1)!}{(j-q)!}<N>C_{j-1}  
\end{equation}
which in turn connects the widely used factorial moments $F_q$  to the combinants through the following recurrence relation
\begin{equation}
  K_q  = F_q - \sum_{i=1}^{q-1}\binom{q-1}{i-1}K_{q-i}F_i
\end{equation}
A comprehensive examination of the statistical properties of modified combinants for various forms of multiplicity distributions and their relationship with entities like the factorial moments can be found in \cite{PhysRevD.105.054003}. It has been shown that the information obtained from the $P(n)$ distribution is limited, and models that describe the multiplicity distribution often fail to reproduce the structure of their modified combinants in many collision systems. This appears to happen due to an inaccurate statistical description of the particle production mechanism.

\vspace{5pt}
{\bf Collisions \& modified combinants --}\label{sec:ModColl} 
The shape of $\left<n\right>C_j$ is affected by the dynamics involved in the collision under study. To demonstrate this, we compare the $\left<n\right>C_j$ of charged particles from non-single diffractive (NSD) proton-proton (pp) collisions with that of inclusive pp collisions at $\sqrt{s} = 900$ GeV, using data from the ALICE experiment at LHC \cite{ALICE:2017pcy}. The majority of hadrons in pp collisions are produced from inelastic non-diffractive scatterings, which occur due to the exchange of color charge. In Regge theory, diffractive events happen when the Pomeron interacts with the proton and produces a system of particles\cite{collins_1977}. Single-diffractive events occur when only one proton dissociates. The type of events is selected using the V0-A and V0-C detectors placed on opposite sides of the ALICE experimental apparatus close to the beam pipe\cite{ALICE_2008}. The INEL (inelastic events) condition refers to events with at least one interaction recorded in the V0-A and V0-C detectors. The NSD condition, on the other hand, requires that the charged particles are detected in both detectors, effectively removing most of the single-diffractive events \cite{ALICE:2017pcy}. The modified combinants of NSD events were studied in detail in \cite{PhysRevD.99.094045}. As seen in FIG.~\ref{fig1}, the $\left<n\right>C_j$ of NSD events shows an oscillatory pattern, while such behavior is not present in the $\left<n\right>C_j$ of the charged particle distributions from INEL events. The difference in proton dissociation between the two types of collisions is possibly reflected in their $\left<n\right>C_j$. Similarly, the combinants of the multiplicity distributions of charged particles in pp collisions measured by the ATLAS collaboration \cite{Aad2016} with similar trigger conditions to ALICE INEL trigger show a non-oscillatory behavior \cite{Zborovsky2018} (FIG. 8). It shows that the modified combinants derived from the experimentally measured $P(n)$ can provide additional information about the production process under study. This information could be used in situations where one would like to distinguish between two competing descriptions that result in very similar multiplicity distributions. In this spirit, we will use the modified combinants to analyze models of deuteron synthesis in nuclear collisions in the following sections. 

\begin{figure*}[t!]
\includegraphics[width=\textwidth,height=0.5\textwidth]{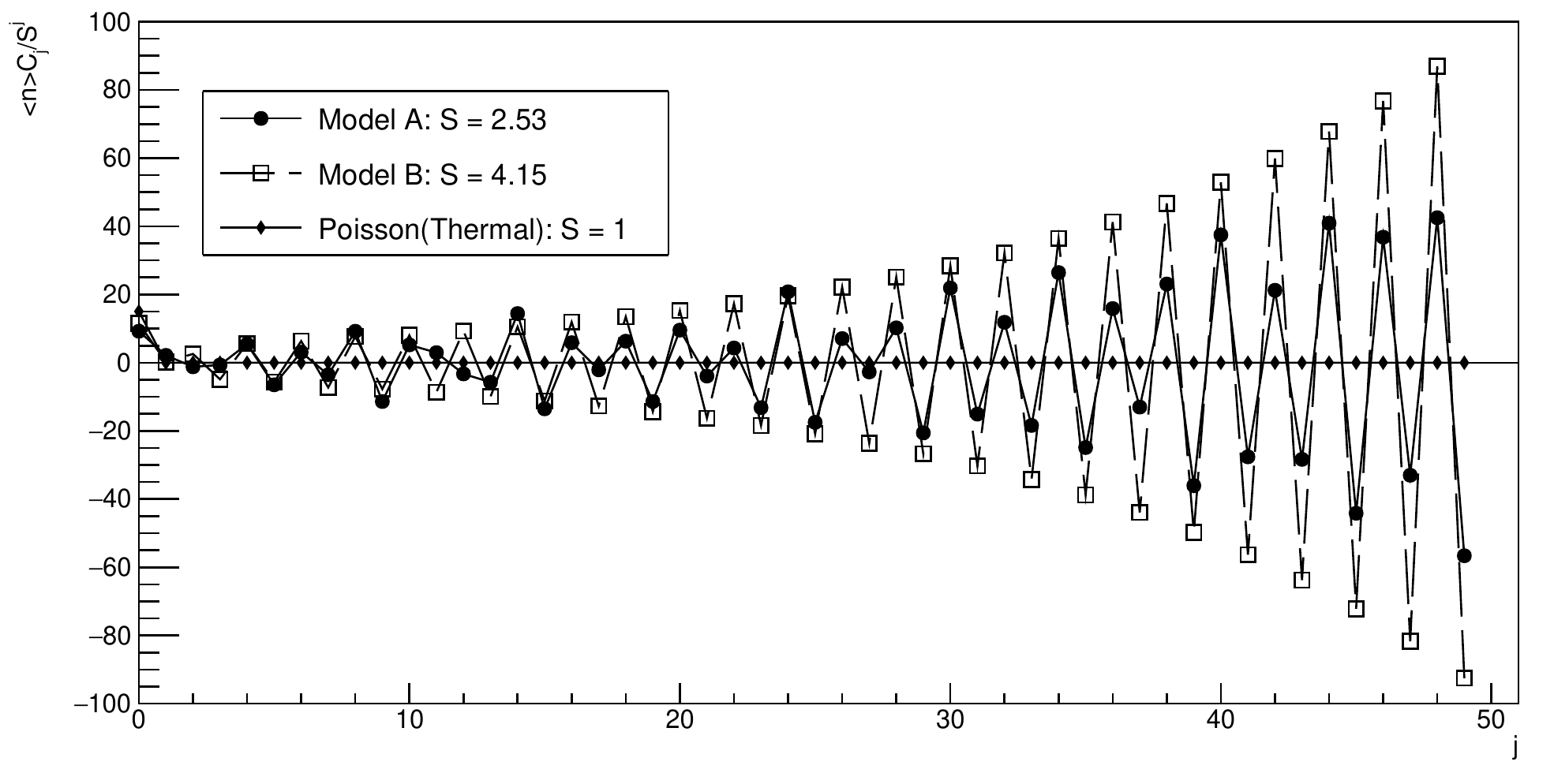}
 \caption{Modified combinants of deuteron multiplicity distribution obtained from the three models of deuteron synthesis mentioned in the text and discussed in \cite{PhysRevC.93.054906}. The solid and dotted lines are to lead the eye.}
 \label{fig2}
\end{figure*}

\vspace{5pt}
{\bf Modified combinants of thermal and coalescence deuterons 
 -}\label{sec:Mod.Deut}
 Several models have been proposed to explain deuteron synthesis in high-energy collisions \cite{VOVCHENKO2018171, SUN2019132, Andronic_2019, Andronic2018}. Deuterons, being composite particles with a small binding energy should be facing difficulty in achieving thermalization through scattering in a QGP-like medium. Nevertheless, the thermal model satisfactorily describes the experimentally observed deuteron yield in various collision systems. There exist different variations of coalescence mechanism towards deuteron formation as well. Here we examine two distinct models of coalescence and a thermal model. The coalescence models we analyze here are described in \cite{PhysRevC.93.054906} and vary in the correlation between proton and neutron number fluctuations during the deuteron formation. In the first model (Model-A), deuterons are formed after kinetic freeze-out in each collision with a production probability, $\lambda_d$, proportional to the square of the initial number of protons, $n_i$. This implies that the production of neutrons is directly correlated with that of protons in the same collision. As a result, the deuteron multiplicity for a given number of protons, $n_i$, is given by:
\begin{equation}
    P_A(n_d|n_i) = (Bn_i^2)^{n_d}\frac{e^{-Bn_i^2}}{n_d!}
\end{equation}
where the coalescence parameter $B$, is the proportionality constant between $\lambda_d$ and $n_i^2$. Summing over an initial Poissonian proton number distribution $P(n_i)$ results in a $P_A(n_d)$ given as:
\begin{equation}
\label{modelA}
P_A(n_d) = \sum_{n_i\geq n_d}(Bn_i^2)^{n_d}\frac{e^{-Bn_i^2}}{n_d!}P(n_i)
\end{equation}
In the second model (Model-B), the proton and neutron numbers fluctuate independently. In this scenario, $\lambda_d$ is proportional to the product of proton and neutron multiplicities, $n_i$ and $n_j$, respectively. The constant of proportionality, $B$, in both models depends only on the collision energy. The number of deuterons for a given number of $n_i$ and $n_j$ is then given by:
\begin{equation}
P_A(n_d|n_i) = (Bn_in_j)^{n_d}\frac{e^{-Bn_in_j}}{n_d!}
\end{equation}
With the $n_j$ and $n_i$ fluctuating according to a Poisson distribution with the same mean, we have the deuteron multiplicity in model B as follows
\begin{equation}
\label{modelB}
 P_B(n_d) = \sum_{n_i,n_j\geq n_d}(Bn_in_j)^{n_d}\frac{e^{-Bn_in_j}}{n_d!}P(n_i)P(n_j)
\end{equation}
The two models of coalescence and the thermal production scenario were studied for Au-Au collisions at $\sqrt{s_{NN}} = 2.6$ GeV in terms of multiplicity distribution and its moments, and the yield ratio of deuterons to protons in \cite{PhysRevC.93.054906}. However the ALICE measurement of  negative Pearson correlation between antiproton and antideuteron do not support Model-A \cite{https://doi.org/10.48550/arxiv.2204.10166}. It is worth mentioning that the calculations in \cite{PhysRevC.93.054906} were performed in a scenario where anti-proton production can be safely neglected, but this is not the case at the top RHIC and LHC energies. Although that remains the case, we intend to qualitatively demonstrate how these three scenarios of deuteron synthesis make impressions in their modified combinants for the consideration of future experimental and phenomenological studies.
\par
The modified combinants are constructed for the thermal model $P(n)$ which follows a Poisson distribution and two scenarios of nucleon coalescence described by Eqs.~\ref{modelA} and \ref{modelB} \footnote{The values of $P(n_d)$ are obtained from a calculation similar to that in \cite{PhysRevC.93.054906} with the parameter values  
inspired from the \href{https://indico.cern.ch/event/592683/contributions/2393793/}{presentation} of Boris Tomasik at Zimanyi School(2016), Budapest. The ratios of the moments for Model A are $\sigma^2/\left<n_d\right> = 1.61142$, $S\sigma = 2.21540$, and $\kappa\sigma^2 = 6.61484$, while for Model B, they are $\sigma^2/\left<n_d\right> = 1.31173$, $S\sigma = 1.63349$, and $\kappa\sigma^2 = 3.33974$. In the case of a Poisson distribution, all three values are equal to unity. The mean for all three distributions is $\left<n_d\right> = 15.05$, where $\sigma^2$ is the variance, S is the skewness, and $\kappa$ is the kurtosis.}. We use Eq.\ref{Comb} to calculate the modified combinants of the deuteron multiplicities. The results are presented in FIG.\ref{fig2}. It can be observed that the pattern and magnitude of the modified combinants of the Poisson distribution, which represents a thermal model of deuteron production, is significantly different from those of deuterons formed through the two coalescence scenarios. Please note that the values of $\left<n\right>C_j$ in FIG.~\ref{fig2} have been scaled for better visualization. The amplitude of the $\left<n\right>C_j$ oscillations for the Poisson distribution is consistent with zero for all $j>0$. In the case of model-A, with fully correlated neutron and proton fluctuations, the period of $\left<n\right>C_j$ oscillations remains two. For model-B, which involves coalescence with independent neutron and proton fluctuations, slight variation in the period of oscillation is observed for small values of $n$. The amplitude and pattern of $\left<n\right>C_j$ oscillations in this case are significantly different from those in the other two scenarios. The amplitude of $\left<n\right>C_j$ oscillations for model-B is more intense compared to that of model-A. Thus we conclude that the process of deuteron production leaves distinct and observable signatures in the modified combinants of the deuteron multiplicity distribution.
\vspace{5pt} 

{\bf Summary and Outlook --}\label{sec:Sum} 
In the previous sections, the method of combinants and the challenges in deuteron synthesis in nuclear collisions were investigated. It was found that the modified combinants of charged particles in inelastic and non-single diffractive proton-proton collisions at the LHC exhibit different behaviors. Deuteron synthesis in nuclear collisions was explored in the context of these modified combinants. The results indicate that the mechanism of deuteron production leaves substantial signatures on the modified combinants of multiplicity distribution, with unique patterns and amplitudes of oscillation in $\left<n\right>C_j$. Based on these findings, we propose to construct the experimental $P(n)$ distribution of deuterons and other light nuclei, along with their modified combinants, in high-energy hadron-hadron and nucleus-nucleus collisions as a function of centrality at RHIC and LHC energies.
\par
From a theoretical perspective, the generating function of the Poisson distribution is given by
\begin{equation}
G(z) = exp(\lambda(z-1))
\end{equation}
where $\lambda$ is the average number of events. The combinants $C^{*}$ are defined as
\begin{equation}
C^{*}_j = \frac{1}{j!}\frac{d^{j}}{dz^{j}}\ln G(z) \Big|_{z=0} = \lambda^j\delta_{1j}
\end{equation}
where $\delta_{1j}$ is the Kronecker delta. The modified combinants are related to the combinants as
\begin{equation}
C_{j} = \frac{j+1}{\left<n\right>}C^{*}_{j+1}
\end{equation}
Therefore, for the Poisson distribution,
\begin{equation}
C_{j>0} = \delta_{0j}
\end{equation}
This means that for all $j>0$, the Poissonian modified combinants exhibit no oscillation \cite{doi:10.1142/S0217751X2150072X}. Thus the presence of oscillation in the experimentally measured $\left<n\right>C_j$ may be seen as a departure from thermal Poisson-like production of deuterons and other nuclei, making modified combinants a potentially valuable tool for studying and possibly falsifying the models of nuclei synthesis at the colliders. A theoretical model capable of predicting both the experimentally observed $P(n)$ and the resulting $\left<n\right>C_j$ may offer a more credible explanation of deuteron and other nuclei synthesis in hadronic and nuclear interactions.
\par
Although the modified combinants are more effective at low multiplicities\cite{PhysRevD.105.054003}, there are experimental challenges in accurately measuring $P(n)$ for small $n$, particularly $P(0)$, due to reduced particle selection efficiencies of high-energy detectors at low multiplicities (refer to FIG.2 in \cite{Aamodt2010}). A method for overcoming a similar issue for combinants-based analysis was proposed in \cite{PhysRevD.105.054003} in the context of particles within a QCD jet where the $P(0)$ was unavailable. For the cases where experimental data is unreliable for $n < Q$ the following procedure may be applied. The Eq.~\ref{eq2} can be rearranged to obtain
\begin{equation}\label{eq13}
g(n) = (n+1)\frac{P(n+1)}{P(n)}
\end{equation}
Thus $g(n)$ can be constructed from the reliable values of $P(n \geq Q)$ by computing $g(i) = (i+1) P(i+1)/P(i)$ for $i$ running from $Q$ to $n_{max}$, which is the maximum value of $n$ in the experimental $P(n)$ distribution. By fitting the obtained $g(n \geq Q)$ versus $n$ down to $n=Q$ with a function $f(n,\lambda_i)$ (where $\lambda_i$ are the fit parameters) and subsequently extrapolating the fit results to $n < Q $, $g(n < Q)$, the $P(n < Q)$ can be estimated. This scheme can potentially be applied to experimental data with unreliable $P(n)$ values for deuterons (or other nuclei) at low multiplicities by choosing an appropriate form for $f(n,\lambda_i)$. In the method of modified combinants we also have the benefit of not needing to know the probabilities themselves. They follow directly from the unnormalized topological cross sections since they involve only ratios of finite number of probabilities \cite{HEGYI1993642}.
\par
The proposed approach offers the possibility to place stronger constraints on the parameters of models for light nuclei synthesis in hadronic and nuclear collisions. On an qualitative note, the correlation between nucleons in coalescence seems to impose constraints on the growth of the amplitude of $\left<n\right>C_j$ of the produced nuclei as a function of $j$. Hence, by examining the pattern of $\left<n\right>C_j$ oscillations and varying the parameters of interest, one may determine the strength of correlation between the neutron and proton numbers in a coalescence scenario. The existence of oscillations in the experimentally measured $\left<n\right>C_j$ can potentially challenge the hypothesis of their thermal Poissonian-like production. Furthermore, a comparison of the shape of modified combinants of nuclei in heavy-ion collisions with those in the hadronic collisions may shed light on possible differences in their production mechanism in the two colliding systems. Applying the method of combinants separately for finding the correct mechanism of deuteron and anti deuteron synthesis in ultrarelativistic heavy-ion collisions might be of fundamental importance in understanding their formation, thereby possibly getting information about the origin of matter-antimatter asymmetry in the universe following the big bang. Hence, the study of (anti)nuclei synthesis via the method of modified combinants with experimental data is a valuable tool to advance our understanding about the universe.

\vspace{5pt}

{\bf Acknowledgments --}\label{sec:Ack} 
RRN acknowledges helpful correspondence with Dr. Jan Steinheimer and Dr. Zuzana Paulinyova (Frankfurt Institute for Advanced Studies). 
\bibliography{references}
\end{document}